\documentclass[12pt]{article} %,a4paper
\usepackage{graphicx}
\usepackage{amsmath}
\usepackage{amssymb}
\allowdisplaybreaks

\usepackage[utf8]{inputenc}
\usepackage{utopia}
\usepackage[utopia]{mathdesign}
\usepackage{eucal}
 \usepackage{cite}
 \usepackage{hyperref}
\usepackage[height=8.85in,width=6.45in]{geometry}
\usepackage{tikz}
\tikzset{>=latex}
\tikzset{baseline=(A.base)}
\tikzset{every picture/.append style={scale=1.1}}
\tikzset{flavor/.style={draw}}
\tikzset{gauge/.style={draw,circle,inner sep=2pt}}

\usepackage{enumitem}
\setlist{noitemsep}

\numberwithin{equation}{section}
\def\Nequals#1{$\mathcal{N}{=}#1$}

\newcommand{\bR}{\mathbb{R}}
\newcommand{\bZ}{\mathbb{Z}}

\def\ket#1{\left| #1 \right\rangle}

\def\node#1{\overset{#1}{\circ}}
\def\blacknode#1{\overset{#1}{\bullet}}
\def\ver{\overset{\displaystyle\circ}{\scriptstyle\vert}}

\def\U{\mathrm{U}}
\def\SU{\mathrm{SU}}

\def\Sp{\mathrm{Sp}}
\def\SO{\mathrm{SO}}

\def\SU{\mathrm{SU}}

\let\oldmid\mid
\def\mid{\,\oldmid\,}

\def\tr{\mathop{\mathrm{tr}}\nolimits}
\def\diag{\mathop{\mathrm{diag}}}

\begin{document}

\begin{titlepage}

\begin{flushright}
IPMU-15-0002\\
UT-15-01\\
\end{flushright}

\vskip 2cm

\begin{center}
{\Large \bfseries
Instanton operators and symmetry enhancement \\[1em]
in 5d supersymmetric gauge theories
}

\vskip 1.5cm
 Yuji Tachikawa
\vskip 1.0cm

\begin{tabular}{ll}
  & Department of Physics, Faculty of Science, \\
& University of Tokyo,  Bunkyo-ku, Tokyo 133-0033, Japan, and\\
  & Institute for the Physics and Mathematics of the Universe, \\
& University of Tokyo,  Kashiwa, Chiba 277-8583, Japan\\
\end{tabular}

\vskip 1cm

\textbf{Abstract}

\end{center}

\medskip
\noindent 
Supersymmetric gauge theories in five dimensions often exhibit less symmetry than the ultraviolet fixed points from which they flow. The fixed points might have larger flavor symmetry or they might even be secretly six-dimensional theories on $S^1$.  
Here we provide a simple criterion when such symmetry enhancement in the ultraviolet should occur, by a direct study of the fermionic zero modes around one-instanton operators.

\end{titlepage}

\setcounter{tocdepth}{2}
\tableofcontents

%%%%%%%%%%%%%%%%%%%%%%%%%%%%%%%%%%%%%%%%%%%%%

\section{Introduction}
In four dimensions and in lower dimensions, we often start from a Lagrangian gauge theory in the ultraviolet (UV) and study the behavior of the system in the infrared (IR).
In five dimensions (5d), Lagrangian gauge theories are always non-renormalizable, and one instead needs to look for nontrivial ultraviolet superconformal field theories (SCFTs)  from which they flow out.
With supersymmetry, such a study is indeed possible, as demonstrated in \cite{Seiberg:1996bd,Morrison:1996xf,Douglas:1996xp,Intriligator:1997pq}, by combining  field-theoretical analyses and embedding into string theory. 

There and in other works, it was found that the UV fixed point can have enhanced symmetry: the instanton number symmetry sometimes enhances to a non-Abelian flavor symmetry, and sometimes enhances to the Kaluza-Klein mode number of a six-dimensional theory on $S^1$. 
For example, the UV SCFT for  \Nequals{1} $\SU(2)$ theory with $N_f\le 7$ flavors has $E_{N_f+1}$ flavor symmetry; with $N_f=8$ flavors, the UV fixed point is instead a 6d \Nequals{(1,0)} theory with $E_8$ symmetry compactified on $S^1$ \cite{Ganor:1996pc}. 
For $\SU(N)$ theory, the flavor symmetry enhancement only occurs for some specific choice of the number of the flavors and of the Chern-Simons levels. \Nequals{2} theory with gauge group $G$ in 5d, in contrast, comes from some 6d \Nequals{(2,0)} theory  compactified on $S^1$, as originally found in the context of string duality \cite{Witten:1995zh,Strominger:1995ac}.

These results were soon extended to include more models, using webs of five-branes, in e.g.~\cite{Aharony:1997ju,Aharony:1997bh,DeWolfe:1999hj}. More recently, various sophisticated  techniques such as supersymmetric localizations, Nekrasov partition functions, and refined topological strings have been applied to the analysis of these 5d systems. The symmetry enhancement of the models mentioned above has been successfully confirmed by these methods, and  even more diverse models are  being explored, see e.g.~\cite{Kim:2012gu,Iqbal:2012xm,Bashkirov:2012re,Bergman:2013koa,Bergman:2013ala,Bao:2013pwa,Hayashi:2013qwa,Taki:2013vka,Bergman:2013aca,Aganagic:2014oia,Taki:2014pba,Hwang:2014uwa,Zafrir:2014ywa,Hayashi:2014wfa,Bergman:2014kza,Hayashi:2014hfa,Mitev:2014jza}. 

These results are impressive, but the techniques used are rather unwieldy.\footnote{One of the trickiest aspects is the need to remove spurious contributions from the so-called ``$\U(1)$ parts'' in the instanton counting method and from the parallel legs in the toric diagram in the topological string method.}
In this paper, we describe a simpler method to identify the symmetry enhancement of a given 5d gauge theory, assuming that it has an ultraviolet completion either in 5d or 6d. Although heuristic, this method tells us what will be the enhanced flavor symmetry and whether the ultraviolet completion is a 5d SCFT or  a 6d SCFT on $S^1$.

We do this by identifying the supermultiplet of broken symmetry currents\footnote{The importance and the special property of the current supermultiplets in 5d SCFTs were emphasized in \cite{Bashkirov:2012re}. The analysis presented here is strongly influenced by that paper. } by studying instanton operators, that introduce non-zero instanton number on a small $S^4$ surrounding a point.\footnote{Recently there appeared a paper \cite{Lambert:2014jna} where instanton operators of 5d \Nequals2 gauge theories were also studied.  There the emphasis was on the multi-point functions of instanton operators. In this paper we just consider a single instanton operator, and the multiplet it forms under the flavor symmetry and the supersymmetry.
In another recent paper \cite{Rodriguez-Gomez:2015xwa}, a different type of instanton operators was studied, where an external $\Sp(1)_R$ background was introduced on $S^4$ to have manifest supersymmetry at the classical level. Here we do not introduce such additional backgrounds.} When the instanton number is one, the structure of the moduli space and the zero modes is particularly simple, allowing us to find the broken symmetry currents rather directly.  The spirit of the analysis will be  close to the analysis of monopole operators of 3d supersymmetric gauge theories in \cite{Gaiotto:2008ak}, except that we need to identify an operator in the IR that would come from a UV current operator, rather than vice versa. 

The rest of the paper is organized as follows. In Sec.~\ref{sec:preliminaries}, we collect basic facts on the supersymmetry and on instantons on $S^4$ that we use. 
In Sec.~\ref{sec:SU2}, we analyze  \Nequals1 $\SU(2)$ gauge theories with $N_f\le 8$ flavors and  \Nequals2 $\SU(2)$ theory. The effect of the discrete theta angle will also be discussed.
In Sec.~\ref{sec:SUN}, we extend the discussion to \Nequals1 $\SU(N)$ gauge theories with fundamentals and Chern-Simons terms, and to \Nequals2 $\SU(N)$ theory.
We use the results that will be obtained up to this point in Sec.~\ref{sec:quivers} to study symmetry enhancement in quiver gauge theories made of $\SU$ gauge groups and bifundamentals. We assume that the effective number of flavors of each $\SU(N)$ node is $2N$ and that the Chern-Simons levels are all zero.
We conclude with a discussion in Sec.~\ref{sec:conclusions}.

\section{Preliminaries}\label{sec:preliminaries}
\subsection{Supermultiplet of broken currents}
Suppose that the 5d supersymmetric gauge theory is obtained by a mass deformation of a 5d superconformal theory.  In this setup, supersymmetric mass deformations are always associated to the Cartan part of a possibly non-Abelian conserved current supermultiplet,\footnote{For the details of the superconformal multiplets in 5d and in 6d, see e.g.~\cite{Bhattacharya:2008zy} and references therein.} that contains  the conformal primaries\begin{equation}
\mu^a_{(ij)}, \quad \psi^a_{i\alpha}, \quad J_\mu^a,  \quad M^a \label{5dmul}
\end{equation} with scaling dimension $3$, $3.5$, $4$, $4$ respectively.
Here $a$ is the adjoint index of the flavor symmetry, $i=1,2$ is the index of $\Sp(1)_R$ and $\alpha$ is the spinor index of $\SO(5)$; the symplectic Majorana condition is imposed on $\psi^a_{i\alpha}$. 

The mass deformation is done by adding $\delta L= h_a M^a$ to the Lagrangian, thus breaking some of the flavor symmetry: \begin{equation}
\partial_\mu J_\mu^a \propto f^{ab}{}_{c} h_b M^c.
\end{equation} where $f^{ab}{}_c$ is the structure constant. 
The theory is no longer superconformal, but we can still consider the supermultiplets of operators under the 5d \Nequals1 supersymmetry, modulo spacetime derivatives $\partial_\mu$.  
No comprehensive analysis of the structure of such supermultiplets in 5d is available in the literature, to the knowledge of the author. We can still see that the supermultiplet containing $\mu^a_{(ij)}$ of a broken generator is short, because the superderivative of 
 $\mu^a_{(ij)}$ only contains $\psi^a_{i\alpha}$ which is a doublet under $\Sp(1)_R$.  This fact allows us to identify the broken current supermultiplets clearly in the infrared gauge theory: they are supermultiplets containing the following:
 \begin{equation}
 \mu_{(ij)},\quad \psi_{i\alpha}, \quad J_\mu \label{broken}
\end{equation}
where $\mu_{(ij)}$ is  an $\Sp(1)_R$ triplet scalar $\mu_{(ij)}$, 
$\psi_{i\alpha}$ is a symplectic Majorana $\Sp(1)_R$ doublet fermion $\psi_{i\alpha}$, and $J_\mu$ is a vector.  
The  component $M$ in \eqref{5dmul} is a divergence of $J_\mu$, and therefore is a descendant. 
 
Suppose instead that the 5d   theory is obtained by putting an \Nequals{(1,0)} 6d SCFT on $S^1$, possibly with a nontrivial holonomy for the flavor symmetry;  we consider \Nequals{(2,0)} theories as special cases of \Nequals{(1,0)} theories.  The conserved current supermultiplet in 6d contains \begin{equation}
\mu_{(ij)}, \quad \psi_{i\alpha}, \quad J_A, \label{6dmul}
\end{equation} where $A$ is the 6d vector index and we omitted the adjoint index for brevity.
On $S^1$, we have Kaluza-Klein (KK) modes $J_\mu^{(n)}$ and $M^{(n)} := J_6^{(n)}$ where $n$ is the KK mode number. Then the 6d conservation gives \begin{equation}
\partial_\mu J_\mu^{(n)} \propto \frac{n}{L} M^{(n)}
\end{equation} in 5d, where $L$ is the circumference of $S^1$.  Therefore, we again find the same supermultiplet of broken currents.

\subsection{Instanton operators}
Suppose now that we are given a 5d gauge theory, and further assume that it is a mass deformation of a 5d SCFT or a 6d SCFT on $S^1$.  
Pick a point $p$ in $\bR^5$ and insert there a current operator that is broken by the mass deformation in the former case, and a current operator with non-zero KK mode number in the latter case. Surround $p$ by an $S^4$, on which we have a certain state. %\footnote{Note that this operation is a general property of quantum field theories and can be used in a massive phase. With conformal invariance we can show the one-to-one correspondence between state and operators.}
Let the size of $S^4$ be sufficiently larger than the characteristic length scale of the system set by the mass deformation or the inverse radius of $S^1$.  
Then the state on $S^4$ can be analyzed using the gauge theory.  When the instanton number of the gauge configuration on $S^4$ is nonzero, we call the original operator inserted at $p$ an instanton operator. 

It should be possible to study the supermultiplet structure of instanton operators in detail using supersymmetric  5d Lagrangians on $S^4$ times $\bR$ or $S^1$, using the results in e.g.~\cite{Kim:2012gu,Terashima:2012ra,Imamura:2014ima,Kim:2014kta}. Here we only provide a rather impressionistic analysis of one-instanton operators, i.e.~the instanton operators when the instanton number on $S^4$ is one.   

In the rest of this section we gather known facts on one-instanton moduli spaces and fermion zero modes. 
We will be brief; a comprehensive account on $\SU(N)$ instantons can be found in e.g.~\cite{Dorey:2002ik}. For instantons of general gauge groups, see e.g.~\cite{Bernard:1977nr,Vainshtein:1981wh}.

\paragraph{When the gauge group is $\SU(2)$:}
Any instanton configuration on $S^4$ can be obtained by conformal transformations from one on a flat $\bR^4$.  When the gauge group is $\SU(2)$, the one-instanton moduli space  on  $\bR^4$ has the form \begin{equation}
\bR^4 \times \bR_{>0} \times  S^3/\bZ_2,
\end{equation} where $\bR^4$ parametrizes the position, $\bR_{>0}$ the size, and $S^3/\bZ_2$ the gauge rotation at infinity.  When the instanton is mapped to $S^4$, the first two factors $\bR^4$ and  $\bR_{>0}$ combine to form the ball $B^5$ with a standard hyperbolic metric.\footnote{The hyperbolic ball $B_5$ is also known as the Euclidean AdS$_5$. This fact was used effectively in a series of early works relating AdS/CFT and instantons, see e.g.~\cite{Bianchi:1998nk,Dorey:1998xe,Dorey:1998qh,Dorey:1999pd}. }  
The asymptotic infinity of $\bR^4$ is mapped to a point on $S^4$, therefore the gauge symmetry there should really be gauged. Therefore we lose the last factor $S^3/\bZ_2$, but we should remember that the gauge group $\SU(2)$ is broken to $\bZ_2$. 

A point in $B^5$ very close to a point $x$ in $S^4$ describes an almost point-like instanton localized at $x$. A point at the center of $B^5$ corresponds to the largest possible instanton configuration, that is in fact $\SO(5)$ invariant. This invariant configuration can be identified with the positive-chirality spinor bundle on the round $S^4$. This configuration is also known as Yang's monopole \cite{Yang:1977qv}.

A Weyl fermion of the correct chirality in the doublet of $\SU(2)$ gauge group has just one zero mode on a one-instanton configuration. On $S^4$, this is a singlet of $\SO(5)$ rotational symmetry.  

A  Weyl fermion of the correct chirality in the adjoint of $\SU(2)$ gauge group has four zero modes.
Two are obtained by applying  supertranslations
and the other two by applying special superconformal transformations to the original bosonic configuration. 
When conformally mapped to $S^4$, these four modes transform in the spinor representation of $\SO(5)$. 
They are exactly the modes obtained by applying the 5d supersymmetry to the bosonic instanton configuration. In this sense the instanton configuration breaks all the supersymmetry classically, but this does not mean that the instanton operator is in a generic, long multiplet of supersymmetry, as we will soon see.

\paragraph{When the gauge group is general:}
When the gauge group is a general simple group $G$, any one-instanton configuration is obtained by embedding an $\SU(2)$ one-instanton configuration by a homomorphism $\varphi:\SU(2)\to G$ determined by a long root.  Therefore, the one-instanton moduli space on $\bR^4$ is of the form \begin{equation}
\bR^4 \times \bR_{>0} \times G/H
\end{equation}
where $H$ is the part of $G$ that is unbroken by the $\SU(2)$ embedded by $\varphi$. 
A further quotient $G/(\SU(2)\times H)$ is known as the Wolf space of type $G$.
The form of $H$ is well-known;
here we only note that for $G=\SU(N)$ we have $H=\U(1)\times \SU(N-2)$. 

The conformal transformation to $S^4$  again combines $\bR^4$ and $\bR_{>0}$ to the ball $B^5$, and we lose $G/H$ as before.  The remaining effect is that we have $H$ as the unbroken gauge symmetry.   Analyzing  the fermionic zero modes of an arbitrary representation $R$ of $G$ around this configuration is not any harder than for $\SU(2)$, since the actual gauge configuration is still essentially that of $\SU(2)$. We only have to decompose $R$ under $\SU(2)\times H$, and to utilize  our knowledge for $\SU(2)$.

\section{$\SU(2)$}\label{sec:SU2}
\subsection{Pure \Nequals1 theory}
After these preparations, let us first consider one-instanton operators of pure \Nequals1 $\SU(2)$ gauge theory. As recalled in the previous section, the one-instanton moduli space on $S^4$ is just the ball $B^5$. We consider a state corresponding to the lowest $\SO(5)$-invariant wavefunction. 

Now we need to take fermionic zero modes into account.  The gaugino is a spinor field in five dimensions, that gives two Weyl fermions with the correct chirality in the adjoint of $\SU(2)$ gauge group and in the doublet of $\Sp(1)_R$ when restricted on $S^4$. 
As recalled in the previous section, the zero modes are in the doublet of $\Sp(1)_R$ and in the spinor of $\SO(5)$ rotational symmetry. Let us denote them by $\lambda_{i\alpha}$,
where $i=1,2$ is for $\Sp(1)_R$ and $\alpha=1,2,3,4$ is for $\SO(5)$. 
The symplectic Majorana condition in 5d means that we need to quantize these zero modes into `gamma matrices' satisfying \begin{equation}
\{ \lambda_{i\alpha}, \lambda_{j\beta} \} = \epsilon_{ij} J_{\alpha\beta}.
\end{equation}
The states on which these zero modes act, then, 
can be found by decomposing the Dirac spinor representation of $\SO(8)$ in terms of its subgroup $\Sp(1)_R\times \SO(5)$ such that the vector representation of $\SO(8)$ becomes the doublet times the quartet. We find the following sixteen states: \begin{equation}
\ket{\mu_{(ij)}^+}, \quad \ket{\psi_{i\alpha}^+}, \quad \ket{J_\mu^+},\label{instop}
\end{equation} that form exactly the broken current supermultiplet \eqref{broken} recalled in the last section.
We put the plus signs as superscripts to remind us that they are one-instanton operators.

Let us assume that this gauge theory is a mass deformation of a UV 5d SCFT. 
Then the UV SCFT should simultaneously have both 
the multiplet that contains $J_\mu^+$ in the gauge theory
and the multiplet that becomes the instanton number current \begin{equation}
J_{\mu}^0 \propto \epsilon_{\mu\nu\rho\sigma\upsilon}\tr F_{\nu\rho} F_{\sigma\upsilon}.
\end{equation} 
Now, $J_\mu^+$ has charge 1 under $J_\mu^0$, because $J_\mu^+$ is a one-instanton operator. Therefore, they should form an $\SU(2)$ flavor symmetry current.  This conclusion agrees with the original stringy analysis \cite{Seiberg:1996bd}.

\paragraph{Effect of the discrete theta angle.}
In the analysis so far, we neglected the effect of the discrete theta angle of the $\SU(2)$ theory. The theta angle is associated to $\pi_4(\SU(2))=\bZ_2$, and therefore only takes the values $\theta=0$ or $\pi$.  On a five-manifold of the form $M\times S^1$ such that the $\SU(2)$ configuration on $M$ has instanton number 1 and there is a nontrivial $\bZ_2$ holonomy around $S^1$,  the  theta angle $\theta=\pi$ assigns an additional sign factor $-1$ in the path integral.\footnote{For a derivation, see e.g.~the appendix A of \cite{Tachikawa:2011ch}.}
On $\bR^4\times S^1$, this has an effect that makes the wavefunctions on the one-instanton moduli space  sections of a nontrivial line bundle with holonomy $-1$ on $S^3/\bZ_2$.

In our setup,  the supermultiplet \eqref{instop} is kept when $\theta=0$, but is projected out when $\theta=\pi$. Therefore, there is an enhancement of the instanton number symmetry to $\SU(2)$ when $\theta=0$, but we see  no enhancement when $\theta=\pi$.  This effect of the discrete theta angle matches what was found originally in \cite{Morrison:1996xf} using stringy analysis. 

A caveat here is that in our crude analysis, we can only say that the states considered so far do not give any broken current supermultiplet. It is logically possible that exciting non-zero modes in the one-instanton sector or considering operators with instanton number 2 or larger gives rise to broken current supermultiplet,  enhancing the instanton number symmetry to $\SU(2)$ even with $\theta=\pi$.  A careful study on this point is definitely worthwhile, but is outside of  the scope of this paper. 
The same caveat is also applicable to the rest of the article, but we will not repeat it. 

\subsection{With fundamental flavors}
Next, let us consider \Nequals1 $\SU(2)$ theory with $N_f$ flavors in the doublet. 
Stated differently, we add $2N_f$ half-hypermultiplets in the doublet of $\SU(2)$.
At the classical Lagrangian level, they transform under $\SO(2N_f)$ symmetry. 

In a one-instanton background, they give $2N_f$ fermionic zero modes, that need to be quantized as gamma matrices: \begin{equation}
\{ \Gamma^{a}, \Gamma^{b} \} = \delta^{ab}\label{flavorgamma}
\end{equation} where $a=1,\ldots,2N_f$ is the index of the vector representation of $\SO(2N_f)$. 
They act on the Dirac spinor representation $S_+ \oplus S_-$ of $\SO(2N_f)$.
Tensoring with the result \eqref{instop} of the quantization of the adjoint zero modes,
we find that the one-instanton operator is a broken current supermultiplet in the Dirac spinor of $\SO(2N_f)$. 

Now we need to take the unbroken $\bZ_2$ gauge symmetry into account.  The generator of this $\bZ_2$ symmetry acts by $-1$ on the doublet representation, and therefore by $-1$ on the gamma matrices in \eqref{flavorgamma}. Therefore, this $\bZ_2$ acts as the chirality operator on $S_+\oplus S_-$.   Therefore, depending on the value of the discrete theta angle being $0$ or $\pi$, we keep only one-instanton operators in $S_+$ or $S_-$. These two choices are related by the parity operation of flavor $\mathrm{O}(2N_f)$ and therefore equivalent. 
In conclusion, we found  the following current multiplets: \begin{itemize}
\item the conserved $\SO(2N_f)$ currents $J^{[ab]}_\mu$,
\item the conserved instanton number current $J^0_\mu$,
\item the broken currents coming from one-instanton operators $J^{+,A}_\mu$ where $A$ is the chiral spinor index of $\SO(2N_f)$.
\end{itemize}

Take $1\le N_f\le 7$.  If we assume that the gauge theory is a mass deformation of a UV fixed point, we see that the currents listed above need to combine to give the flavor symmetry $E_{N_f+1}$. This can be seen by attaching an additional node, representing a Cartan element for $J^0_\mu$, to the node of the Dynkin diagram of $\SO(2N_f)$ that gives the chiral spinor representation. 
 For example, we have \begin{equation}
 \blacknode{}-\node{}-\node{\ver}-\node{}-\node{}-\node{}-\node{}
 \end{equation} when $N_f=7$, where the black node is for the instanton number current.\footnote{Strictly speaking, this procedure is not unique when a theory with very small $N_f$ is considered alone. As a simpler example of this issue, suppose we know we have an $\SU(2)$ currents, an instanton number current, and a broken current coming from one-instanton operators in the doublet of $\SU(2)$.  Then we have two choices, either an $\SU(3)$ corresponding to $\blacknode{}-\node{}$ or an $\Sp(2)$ corresponding to $\blacknode{}\Leftarrow \node{}$. They can only be distinguished by studying the two-instanton operators. 
In our case, however, we can just study the $N_f=7$ case and then apply the mass deformation to make $N_f$ smaller, to conclude that the flavor symmetry is always $E_{N_f+1}$. }  
 This enhancement pattern agrees with what was found originally in \cite{Seiberg:1996bd}.

Let us boldly take $N_f=8$. We now have the combined Dynkin diagram \begin{equation}
 \blacknode{}-\node{}-\node{\ver}-\node{}-\node{}-\node{}-\node{}-\node{}
\end{equation} which is known as $E_9$ or $\hat E_8$. 
If we assume that the gauge theory is an outcome of a massive deformation of some UV completion,
the UV completion needs to have an \emph{affine} $E_8$ flavor symmetry as a 5d theory. 
Stated differently, this means that the UV completion needs to be a 6d theory with $E_8$ flavor symmetry. This conclusion matches with what was found in \cite{Ganor:1996pc}.

We see a problem when $N_f>8$: the combined Dynkin diagram defines a hyperbolic Kac-Moody algebra, whose number of roots grows exponentially. It is therefore unlikely that there is a UV completion when $N_f>8$. Again, this matches the outcome of a different analysis in \cite{Seiberg:1996bd}.

Before proceeding, we note here that it has been well-known that the instanton particles in this gauge theory give rise to spinors of $\SO(2N_f)$ flavor symmetry. The only slightly new point in this section is that when studied in the context of instanton operators, they are indeed part of broken current multiplets given by \eqref{instop}.

\subsection{\Nequals2 theory}
As a final example of $\SU(2)$ gauge theory, let us consider \Nequals2 gauge theory. 
On the one-instanton background on $S^4$, we have four Weyl fermions in the adjoint of $\SU(2)$, and the zero modes can be denoted by $\lambda_{i\alpha}$ where $i=1,2,3,4$ is now for $\Sp(2)_R$ and $\alpha=1,2,3,4$ is for the spacetime $\SO(5)$. Again, the symplectic Majorana condition in 5d means that they become `gamma matrices' with the commutation relation \begin{equation}
\{\lambda_{i\alpha},\lambda_{j\beta}\}
=J_{ij}J_{\alpha\beta}.
\end{equation}
These gamma matrices act on the following one-instanton states on $S^4$: \begin{equation}
\ket{\mu^+_{(ab)}},\quad
\ket{\psi^+_{ai\ \alpha}},\quad
\ket{J^+_{[ab]\ \mu }},\quad
\ket{Q^+_{i \ \mu\alpha }},\quad
\ket{T^+_{(\mu\nu)}},\quad
\ket{X^+_{a \ [\mu\nu] }},
\label{2,0}
\end{equation}
where $a=1,2,3,4,5$ is the vector index for $\Sp(2)_R=\SO(5)_R$.
The gamma-tracelessness condition on $\psi^+$, $Q^+$
and the tracelessness condition on $\mu^+$, $T^+$ need to be imposed. 

When the discrete theta angle is zero, these operators are all kept, and the structure of the operators
is exactly that of KK modes of the 6d \Nequals2 energy-momentum supermultiplet.  
For example, 
the currents $J^+_{[ab]\ \mu}$ in the adjoint of $\Sp(2)_R$ suggest that
the $\Sp(2)_R$ symmetry of the 5d gauge theory enhances to the affine $\Sp(2)_R$,
and the symmetric traceless $T_{(\mu\nu)}^+$ is the KK mode of the 6d energy-momentum tensor.
This is as it should be, since the $S^1$ compactification of \Nequals{(2,0)} theory of type $\SU(2)$ on $S^1$ is described by \Nequals2 $\SU(2)$ gauge theory in 5d.  The relation of the 5d \Nequals2 theory and the 6d \Nequals{(2,0)} theory on $S^1$ has been extensively studied, see e.g.~\cite{Douglas:2010iu,Lambert:2010iw}.

When the discrete theta angle is $\pi$, these operators are all projected out.  In this case too, the gauge theory is the IR description of \Nequals{(2,0)} theory of type $\SU(3)$ on $S^1$ with an outer-automorphism $\bZ_2$ twist around it \cite{Tachikawa:2011ch}. We expect that the operators with the same structure to arise in a sector with higher instanton number, but to check it is outside of the scope of this paper.

\section{$\SU(N)$}\label{sec:SUN}
\subsection{Pure \Nequals1 theory}
Now let us move on to $\SU(N)$ gauge theories. Our first example is the pure $\SU(N)$ theory. 
As recalled in Sec.~\ref{sec:preliminaries}, in one-instanton configurations on $S^4$, the gauge fields take values in $\SU(2)\subset \SU(N)$. The unbroken subgroup is $\U(1)\times \SU(N-2)$.
We take the generator of the $\U(1)$ part to be \begin{equation}
\diag(N-2,N-2,-2,-2,\ldots,-2).
\end{equation}
In this normalization, when the Chern-Simons level is $\kappa$, the one-instanton configuration has the $\U(1)$ charge $(N-2)\kappa$.

An adjoint Weyl fermion of $\SU(N)$ on this background decomposes into 
an $\SU(2)$ adjoint Weyl fermion, and $N-2$ Weyl fermions in the doublet of $\SU(2)$ in the fundamental of $\SU(N-2)$ and with $\U(1)$ charge $N$, and Weyl fermions that are neutral.
The gaugino in 5d gives two adjoint Weyl fermions of $\SU(N)$ on $S^4$.
The $\SU(2)$ adjoint part gives the same broken current supermultiplet \eqref{instop},
and we need to take additional $2\times (N-2)$ zero modes coming from the $\SU(2)$ doublet into account. 

By quantizing them, we have fermionic creation operators $B_{ia}$ where $i=1,2$ is now for $\Sp(1)_R$ and $a=1,\ldots,(N-2)$ is for $\SU(N-2)$. The $\U(1)$ charge of $B_{ia}$ is $N$.
 The $\SU(N-2)$ invariant states are then \begin{equation}
\ket 0,\quad
\epsilon^{a_1\cdots a_{N-2}}
B_{i_1 a_1} B_{i_2 a_2} \cdots B_{i_{N-2}a_{N-2}}
\ket 0,\quad
(B_{ia})^{\wedge 2(N-2)}\ket 0,\label{sunadj}
\end{equation} with $\U(1)$ charge $-(N-2)N$, $0$, $+(N-2)N$, respectively. 

When $\kappa$ is neither 0 nor $\pm N$, all states are projected out, due to the nonzero $\U(1)$ gauge charge.  When $\kappa$ is $\pm N$,  one singlet state is kept, and we have a broken current supermultiplet \eqref{instop}. Therefore we expect the enhancement of the instanton number symmetry to $\SU(2)$.  This enhancement was recently discussed in \cite{Bergman:2013aca}.

When $\kappa$ is $0$,  the one-instanton operators are the tensor product of the broken current supermultiplet \eqref{instop} times $\epsilon^{a_1\cdots a_{N-2}}
B_{i_1 a_1} B_{i_2 a_2} \cdots B_{i_{N-2}a_{N-2}}
\ket 0$, which transforms in the $N-1$ dimensional irreducible representation of $\Sp(1)_R$.  This is a short supermultiplet, but does not correspond to a broken flavor symmetry. 

\subsection{With fundamental flavors}
Our next example is  $\SU(N)$ theory with $N_f$ hypermultiplets in the fundamental representation.  
On one-instanton configurations on $S^4$, each flavor decomposes into a pair of $\SU(2)$ doublets and a number of neutrals; they all have $\U(1)$ charge $N-2$.  Therefore, we have additional fermionic creation operators $C_{s}$, $s=1,\ldots,N_f$, of $\U(1)$ charge $N-2$.  
They act on the states of the form \begin{equation}
C_{s_1}C_{s_2}\cdots C_{s_k} \ket 0\label{sunfun}
\end{equation} for $k=0,\ldots,N_f$, with $\U(1)$ charge $(N-2)(k-N_f/2)$. 

Tensoring \eqref{instop}, \eqref{sunadj} and \eqref{sunfun} and imposing the $\U(1)$ gauge neutrality condition, we see that a broken symmetry supermultiplet survives when we have  \begin{equation}
\kappa \pm N + (k - N_f/2)=0.\label{?}
\end{equation}

Now, in \cite{Intriligator:1997pq} it was shown that we need $|\kappa| \le N-N_f/2$ to have a 5d UV SCFT behind the gauge theory. Therefore $|\kappa \pm N| \ge N_f/2$. 
We also trivially have $|k-N_f/2| \le N_f/2$. Therefore, the equality \eqref{?} can only be satisfied when $\pm\kappa=N-N_f/2$.

When $\pm\kappa=N-N_f/2$,
the surviving broken current supermultiplet comes from $k=0$ or $k=N_f$ in \eqref{sunfun}. They have instanton number one and baryonic charge $\pm N_f/2$.
Therefore, when $\kappa = N-N_f/2$, one $\U(1)$ enhances to $\SU(2)$,
and when $-\kappa=N-N_f/2$, another $\U(1)$ enhances to $\SU(2)$. 
When $\kappa=N-N_f/2=0$, 
 the combination $I \pm B / N$ 
 of the instanton charge $I$ and the baryonic charge $B$ 
 both enhance to $\SU(2)_\pm$, making the UV flavor symmetry to be $\SU(N_f)\times \SU(2)_+\times \SU(2)_-$.
This enhancement pattern was found in \cite{Bergman:2013aca,Mitev:2014jza}.

\subsection{\Nequals2 theory}
Let us next consider \Nequals2 $\SU(N)$ theory. The Chern-Simons level $\kappa$ is automatically zero.  Again, the adjoint Weyl fermions of $\SU(N)$ decompose into those that are adjoint of $\SU(2)$ and those that are doublets of $\SU(2)$. The zero modes of the former generate the states \eqref{2,0}.

The zero modes of the latter give us fermionic creation operators $B_{ia}$, where $i=1,2,3,4$ is for $\Sp(2)_R$ and $a=1,\ldots,(N-2)$ is for gauge $\SU(N-2)$. The $\U(1)\times \SU(N-2)$ neutral states have then the form \begin{multline}
\epsilon^{a_1\cdots a_{N-2}}
B_{i_1 a_1} B_{i_2 a_2} \cdots B_{i_{N-2}a_{N-2}}
\epsilon^{b_1\cdots b_{N-2}}
B_{j_1 b_1} B_{j_2 b_2} \cdots B_{j_{N-2}b_{N-2}}
\ket 0 \\
= 
B_{i_1\,1}B_{j_1\,1}
B_{i_2\,2}B_{j_2\,2}
\cdots
B_{i_{N-2}\,N-2}B_{j_{N-2}\,N-2}
\ket 0. \label{many}
\end{multline}

We want to decompose them under the action of $\Sp(4)_R$.  
Let us think $\SU(4)$ acts on the indices $i$ and $j$.
The indices $i_n$ and $j_n$ are antisymmetrized. Combined, $[i_nj_n]$ is a vector of $\SO(6)\simeq \SU(4)$. 
The indices are symmetrized under the combined exchange of $[i_nj_n]$ and $[i_mj_m]$.
Therefore, the states \eqref{many} transform under the $(N-2)$-nd symmetric power of the vector of $\SO(6)$. Decomposing it under $\Sp(4)_R\simeq \SO(5)_R \subset \SO(6)$, we see that the states \eqref{many} are in the representation \begin{equation}
V_{0} \oplus V_{1} \oplus \cdots V_{N-2},\label{aho}
\end{equation}  where $V_{k}$ is the $k$-th symmetric traceless representation of $\SO(5)_R$. 

This structure is precisely what we would expect for the KK modes of \Nequals{(2,0)} theory of type $\SU(N)$ put on $S^1$. 
In general, \Nequals{(2,0)} theory of type $G$ has  short multiplets containing a spacetime symmetric traceless tensor that is in $V_{n-2}$ of $\SO(5)_R$, for each $n$ that gives a generator of  invariant polynomials of  $G$. For  $G=\SU(N)$, $n$ runs from $2$ to $N$, thus giving \eqref{aho} tensored with \eqref{2,0}.

It would be interesting to perform similar computations for \Nequals2 theories for other $G$. For example, when $G=E_6$ we should have $V_0\oplus V_3 \oplus V_4 \oplus V_6 \oplus V_7 \oplus V_{10}$, since the generators of invariant polynomials of $E_6$ have degrees 2, 5, 6, 8, 9,  and 12. 
When $G$ is non-simply-laced, the UV completion is \Nequals{(2,0)} theory of some simply-laced type, with an outer-automorphism twist around $S^1$. This again predicts which $V_n$ should appear among one-instanton operators.  

\section{Quivers}\label{sec:quivers}

In this section, we study the symmetry enhancement in the quiver gauge theory with $\SU$ gauge groups and bifundamental hypermultiplets.  In this note we do not aim comprehensiveness; instead we only treat the case where the effective number of flavors at each node $\SU(N_i)$ is $2N_i$ and the Chern-Simons levels are all zero. 

We use the by-now standard notation where \begin{tikzpicture}
\node[flavor] (A) at (0,0) {$N_1$};
\node[gauge] (B) at (1,0) {$N_2$};
\draw (A)--(B);
\end{tikzpicture} stands for an $\SU(N_1)$ flavor symmetry node 
and an $\SU(N_2)$ gauge symmetry node connected by  a bifundamental hypermultiplet, etc.
We also use a special convention that two flavors of ``$\SU(1)$'', \begin{tikzpicture}
\node[flavor] (A) at (0,0) {$2$};
\node[gauge] (B) at (1,0) {$1$};
\draw (A)--(B);
\end{tikzpicture}, stands for \begin{tikzpicture}
\node[flavor] (A) at (0,0) {$2$};
\node[flavor] (B) at (1,0) {$2$};
\draw (A)--(B);
\end{tikzpicture}. The rationale behind this convention will be explained later in Sec.~\ref{sec:special}.

\subsection{$\SU(2)^2$ theory}
Let us begin our analysis by considering the theory \begin{equation}
\begin{tikzpicture}
\node[flavor] (A) at (0,0) {$2$};
\node[gauge] (B) at (1.5,0) {$2$};
\node[gauge] (C) at (3,0) {$2$};
\node[flavor] (D) at (4.5,0) {$2$};
\draw (A)--(B) node[midway,above] {$Q_0$};
\draw (B)--(C) node[midway,above] {$Q_1$};
\draw (C)--(D) node[midway,above] {$Q_2$};
\end{tikzpicture}\ .
\end{equation}
We denote the gauge groups as $\SU(2)_1\times \SU(2)_2$.
The hypermultiplets $Q_0$, $Q_1$, $Q_2$ have flavor symmetries $\SO(4)_{F0}$, $\SU(2)_{F1}$, $\SO(4)_{F2}$ respectively.  
The gauge group $\SU(2)_1$ effectively has $N_f=4$ flavors, and thus it has $E_{4+1}=\SO(10)$ flavor symmetry when $\SU(2)_2$ is not gauged. 
After gauging, the remaining flavor symmetry is the commutant of $\SU(2)_2$, which is $\SU(4)\times\SU(2)$.   We can summarize this enhancement pattern as \begin{equation}
\node{}\quad\node{}-\blacknode{}-\node{}
\end{equation} where the two white nodes on the left is $\SO(4)_{F0}$, the white node on the right is  $\SU(2)_F$ , and the black node is the contribution from the instanton operator of $\SU(2)_1$.

The same argument can be applied to the $\SU(2)_2$ side, and we conclude that the full flavor symmetry is \begin{equation}
\node{}\quad\node{}-\blacknode{}-\node{}-\blacknode{}-\node{}\quad\node{},
\end{equation}i.e.~$\SU(2)\times \SU(6)\times \SU(2)$.

Let us study one implication of this enhancement.  The adjoint of $\SU(6)$ can be decomposed as follows: \begin{equation}
\left(
\begin{array}{c|c|c}
A & B_1 & C \\
\hline
B_1^\dagger & A' & B_2 \\
\hline
C^\dagger & B_2^\dagger & A''
\end{array}
\right),\label{blocks}
\end{equation} where each symbol stands for a $2\times 2$ block.
The blocks $A$, $A'$ and $A''$ are three $\SU(2)$ flavor symmetries that can be seen in the Lagrangian; $B_i$ comes from  one-instanton operators of the $\SU(2)_i$ gauge group;
and $C$ comes from instanton operators that have instanton number one for both gauge groups $\SU(2)_{1,2}$.  Let us call these last ones $(1,1)$-instanton operators. 
They transform as chiral spinors under $\SO(4)_{F0,F2}$ and are neutral under $\SU(2)_F$.

Let us try to study $(1,1)$-instanton operators directly. The zero modes of gauginos of $\SU(2)_{1,2}$ give two copies of the broken current multiplets \eqref{instop};  those of hypermultiplets $Q_0$ and $Q_2$ give the spinors of $\SO(4)_{F0}$ and $\SO(4)_{F2}$. 

Finally, the hypermultiplet $Q_1$ couples to one-instanton configurations of both $\SU(2)_{1,2}$. 
Therefore this is effectively a triplet  coupled to an $\SU(2)$ one-instanton configuration. It is also a doublet of $\SU(2)_F$. Therefore they give rise to the states \begin{equation}
\ket{\mu_{(ab)}^+}, \quad \ket{\psi_{a\alpha}^+}, \quad \ket{J_\mu^+}.\label{new}
\end{equation} where $a,b=1,2$ is now the index of the doublet of $\SU(2)_F$. 

We therefore need to take the tensor product of two copies of the broken current multiplet \eqref{instop}, the spinors of $\SO(4)_{F0,F2}$, and the multiplet  \eqref{new}. We then need to impose the $\bZ_{2}$ projections for $\SU(2)_{1,2}$. 

At present we do not know enough about the behavior of the tensor product of the supersymmetry multiplets in 5d. Instead, we learn the following by using the knowledge of the flavor symmetry properties of $(1,1)$-instanton operators deduced from the block decomposition \eqref{blocks}:  \emph{the tensor product of two copies of broken current multiplet \eqref{instop} and the multiplet \eqref{new} contains again a unique broken current multiplet}.
Furthermore, \emph{it is $\SU(2)_F$ neutral, and therefore it comes from the factor $J_\mu^+$ in \eqref{new}.}

\subsection{$\SU(N_1)\times \SU(N_2)$ theory}
Next let us discuss more general two-node quivers given by \begin{equation}
\begin{tikzpicture}
\node[flavor] (A) at (0,0) {$N_0$};
\node[gauge] (B) at (1.5,0) {$N_1$};
\node[gauge] (C) at (3,0) {$N_2$};
\node[flavor] (D) at (4.5,0) {$N_3$};
\draw (A)--(B) node[midway,above] {$Q_0$};
\draw (B)--(C) node[midway,above] {$Q_1$};
\draw (C)--(D) node[midway,above] {$Q_2$};
\end{tikzpicture}\ .
\end{equation}
We assume $N_1>2$, $N_2>2$,  $N_0+N_2=2N_1$ and $N_1+N_3=2N_2$. 
We also set both the Chern-Simons levels to be zero.
Let us denote by $\U(1)_{B1}$ and $\U(1)_{B2}$ the baryonic flavor symmetries that assign
charge 1 to a field in the fundamental of $\SU(N_1)$ and $\SU(N_2)$, respectively. 
Let us also denote by $\U(1)_{I_1}$ and $\U(1)_{I_2}$ the instanton number charge of $\SU(N_1)$ and $\SU(N_2)$, respectively. 

We already saw that the combinations $I_{1\pm}:= I_1 \pm B_1/N_1$ 
and $I_{2\pm}:= I_2 \pm B_2/N_2$  are each enhanced to an $\SU(2)$, giving $\SU(2)^4$ flavor symmetry. Let us show that $I_{1+}$ and $I_{2+}$ combine to form an $\SU(3)_+$ and similarly that 
$I_{1-}$ and $I_{2-}$ combine to form an $\SU(3)_-$.

To see this, we need to analyze $(1,1)$-instanton operators in this theory.  
The gauge group $\SU(N_i)$ is broken to $\U(1)_i\times \SU(N_i-2)$.
The gaugino zero modes can be analyzed as before.
The hypermultiplets $Q_0$ and $Q_2$ give doublets of $\SU(2)$ one-instanton configuration;
the hypermultiplet $Q_1$ give similarly a lot of doublets and just one triplet of $\SU(2)$ one-instanton configuration. 

Then, we need to find states that are neutral under the unbroken gauge group from the tensor product of the following contributions:
\begin{enumerate}
\item[1.] From fields that are doublets of the $\SU(2)$ one-instanton configuration, we have
\begin{itemize}
\item[1a.] contributions \eqref{sunadj} from gauginos of $\SU(N_1)$ and $\SU(N_2)$,
\item[1b.] and contributions \eqref{sunfun} from $Q_0$, $Q_1$ and $Q_2$.
\end{itemize}
\item[2.] From fields that are triplets of  the $\SU(2)$ one-instanton configuration, we have
\begin{itemize}
\item[2a.]  a contribution \eqref{new} from $Q_1$,
\item[2b.] and two copies of the broken current multiplet \eqref{instop} from $\SU(N_{1,2})$.
\end{itemize}
\end{enumerate}

From the contributions 1, we find two states that are neutral under the unbroken gauge group $\U(1)_{1}\times \SU(N_1-2)\times \U(1)_2\times \SU(N_2-2)$,
by tensoring  the ground state or the top state of \eqref{sunfun} from the contributions 1a
by the ground state or the top state of \eqref{sunfun} from the contributions 1b. 

From the contributions 2a, we note that the only $\U(1)_1$- and $\U(1)_2$- neutral state is the component $J_\mu$ in \eqref{new}.  Tensoring with the contributions 2b, we find a broken current multiplet, as we found at the end of the last subsection. 

In total, we find at least two gauge-invariant broken current multiplets. The charges under the Lagrangian flavor symmetries can be easily found: both are neutral under $\SU(N_0)$ and $\SU(N_3)$, and the charges under $\U(1)_{Q0}$, $\U(1)_{Q1}$, $\U(1)_{Q2}$
 are \begin{equation}
\pm (\frac{N_0}2,\frac{N_1-N_2}2,-\frac{N_3}2).
\end{equation}

Therefore we have found two broken current multiplets with charges under $(I_{1+},I_{2+};I_{1-},I_{2-})$ given by $(1,1;0,0)$ and $(0,0;1,1)$ respectively. 
We already know that one-instanton operators of $\SU(N_1)$ give broken current multiplets with charges $(1,0;0,0)$ and $(0,0;1,0)$, and similarly those of $\SU(N_2)$ give multiplets with charges $(0,1;0,0)$ and $(0,0;0,1)$.
Therefore we see that the instanton number currents $I_{1+}$, $I_{2+}$ combine to form $\SU(3)_+$, and the currents $I_{1-}$ and $I_{2-}$  combine to form $\SU(3)_-$. 
The total flavor symmetry is therefore at least 
\begin{equation}
\SU(3)_+ \times \SU(3)_- \times \SU(N_0)\times \SU(N_3)\times \U(1).\label{twonodes}
\end{equation}
  The last $\U(1)$ is absent when $N_0$ or $N_3$ is zero. 

\subsection{Some special two-node quivers}\label{sec:special}
We need to analyze separately the cases when one of the gauge group is $\SU(2)$ or ``$\SU(1)$''.
As already mentioned, we use the convention where ``two flavors of $\SU(1)$'' \begin{equation}
\begin{tikzpicture}
\node[flavor] (A) at (0,0) {$2$};
\node[gauge] (B) at (1,0) {$1$};
\draw (A)--(B);
\end{tikzpicture}\ \phantom{.}
\end{equation} stand for \begin{equation}
\begin{tikzpicture}
\node[flavor] (A) at (0,0) {$2$};
\node[flavor] (B) at (1,0) {$2$};
\draw (A)--(B);
\end{tikzpicture}\ .\label{bosh}
\end{equation}  We also apply the same convention where the $\SU(2)$ on the left is gauged. 

From a purely five-dimensional field theory point of view, this is really just a convention, but is useful because ``$\SU(1)$  with two flavors'' shows an enhanced symmetry of 
$\SU(2)\times \SU(2)\times \SU(2)$, just as a special case of $\SU(N)$ with $2N$ flavors with the symmetry enhancement $\SU(2)\times \SU(2)\times \SU(2N)$. 

From a string/M theory point of view, when ``$\SU(1)$ with two flavors'' is engineered, say in the brane web construction, one in fact finds additional hypermultiplets coming from ``point-like $\SU(1)$ instantons'' that naturally give rise to the setup  \eqref{bosh}. This is another rationale for our convention. 

Now, let us couple this ``$\SU(1)$ with two flavors'' to an $\SU(2)$ gauge group to form a two-node quiver: \begin{equation}
\begin{tikzpicture}
\node[flavor] (X) at (-1,0) {$3$};
\node[gauge] (A) at (0,0) {$2$};
\node[gauge] (B) at (1,0) {$1$};
\draw (X)--(A)--(B);
\end{tikzpicture}\ .
\end{equation}  Using our convention, this is just $\SU(2)$ with five flavors that show an enhancement to $E_6$. This contains $\SU(3)_+\times \SU(3)_- \times \SU(3)$, showing the general pattern we found in \eqref{twonodes}.

We already treated \begin{equation}
\begin{tikzpicture}
\node[flavor] (X) at (-1,0) {$2$};
\node[gauge] (A) at (0,0) {$2$};
\node[gauge] (B) at (1,0) {$2$};
\node[flavor] (Y) at (2,0) {$2$};
\draw (X)--(A)--(B)--(Y);
\end{tikzpicture}\ ,
\end{equation} and saw that the symmetry is $\SU(2)\times\SU(6)\times \SU(2)$. 
As $\SU(6)\supset \SU(3)_+\times \SU(3)_-\times \U(1)$, it again shows the general pattern \eqref{twonodes}. 

Finally let us consider \begin{equation}
\begin{tikzpicture}
\node[flavor] (X) at (-1,0) {$6$};
\node[gauge] (A) at (0,0) {$4$};
\node[gauge] (B) at (1,0) {$2$};
\draw (X)--(A)--(B);
\end{tikzpicture}\ .
\end{equation} The $\SU(2)$ theory before coupling to $\SU(4)$ has an enhanced symmetry $\SO(10)$. After coupling to $\SU(4)$ the remaining part is $\SU(2)_+\times \SU(2)_-$, and they are enhanced by the dynamical $\SU(4)$ to $\SU(3)_+\times \SU(3)_-$, again following the general pattern \eqref{twonodes}.

\subsection{Multi-node quivers}
After our preparation on the two-node quivers, it is easy to analyze general multi-node quivers, again with the restriction that each $\SU(N)$ node has effectively $2N$ flavors and zero Chern-Simons terms. 

Consider as an example the quiver \begin{equation}
\begin{tikzpicture}
\node[flavor] (X) at  (-5,0)  {$5$};
\node[gauge] (Y) at  (-4,0)  {$4$};
\node[gauge] (Z) at  (-3,0)  {$3$};
\node[gauge] (B) at  (-2,0)  {$2$};
\node[gauge] (A) at  (-1,0)  {$1$};
\draw (X)--(Y)--(Z)--(B)--(A);
\end{tikzpicture}\ .
\end{equation} Each $\SU(N_i)$ node with $N_i=4,3,2,1$ shows an enhancement of the linear combination of the instanton current and the baryonic current,  $I_{i\pm }= I_i \pm B_i/N_i$ to $\SU(2)_{i\pm}$.  For each neighboring pair of nodes $\SU(N_i)-\SU(N_{j})$, $\SU(2)_{i\pm}$ 
and $\SU(2)_{j\pm}$ enhance to form $\SU(3)_{\pm}$. 
Therefore, in total, we should have $\SU(5)_+\times \SU(5)_-$ from the enhancement of the instanton number symmetry and the baryonic symmetry. Combined with the original flavor symmetry $\SU(5)$ of the leftmost node, we have \begin{equation}
\SU(5)_+\times \SU(5)_-\times \SU(5)
\end{equation} as the enhanced symmetry. We can easily generalize this analysis to an analogous linear quiver with the gauge group $\SU(N{-}1)\times \SU(N{-}2)\times \cdots \times \SU(2)\times$ `` $\SU(1)$'', with bifundamentals between the neighboring gauge nodes and additional $N$ fundamentals for $\SU(N{-}1)$; we see the symmetry $\SU(N)_+\times \SU(N)_- \times \SU(N)$. 
The 5d SCFT is called the 5d $T_N$ theory, and this linear quiver presentation was recently studied in \cite{Hayashi:2013qwa,Aganagic:2014oia,Bergman:2014kza,Hayashi:2014hfa}.

As another example, consider the following quiver:
\begin{equation}
\begin{tikzpicture}
\node[gauge] (A) at  (-5,0)  {$1m$};
\node[gauge] (Y) at  (-4,0)  {$2m$};
\node[gauge] (Z) at  (-3,0)  {$3m$};
\node[gauge] (X) at  (0,1)  {$3m$};
\node[gauge] (B) at  (-2,0)  {$4m$};
\node[gauge] (C) at  (-1,0)  {$5m$};
\node[gauge] (D) at  (0,0)  {$6m$};
\node[gauge] (E) at  (1,0)  {$4m$};
\node[gauge] (F) at  (2,0)  {$2m$};
\draw (X)--(D)--(E)--(F);
\draw (A)--(Y)--(Z)--(B)--(C)--(D);
\end{tikzpicture}\ .
\end{equation}
We have gauge groups $\SU(N_i)$ with $i=1,2,\ldots,9$, with $N_i=k_im$. 
Let us define $I_{i\pm} =I_i \pm B_i/N_i$.
Then, applying exactly the same argument as above, we see that the currents $I_{i+}$ combine to form $(\hat E_8)_+$ and the currents $I_{i-}$ enhance to $(\hat E_8)_-$. 
The total flavor symmetry is not quite their product, however.  The Cartan generator corresponding to a pure KK momentum of $(\hat E_8)_+$ is \begin{equation}
\sum k_i I_{i+} = \sum k_i I_i + \frac1m\sum B_i
\end{equation} but $\sum B_i$ does not act on  the hypermultiplets and therefore is trivial.  Thus we have \begin{equation}
\sum k_i I_{i+}=\sum k_i I_{i-},
\end{equation} meaning that the total flavor symmetry is \begin{equation}
\widehat {E_8\times E_8},
\end{equation} showing that the possible UV completion of this gauge theory is a 6d SCFT with flavor symmetry $E_8\times E_8$ on $S^1$.  This is as expected: $m$ M5-branes on the ALE  singularity of type $E_8$ gives a 6d \Nequals{(1,0)} SCFT in the infrared, with $E_8\times E_8$ flavor symmetry. Compactifying it on $S^1$ and reducing it to type IIA, we have $m$ D4-branes probing the ALE singularity of type $E_8$. Using the standard technique \cite{Douglas:1996sw}, we find the quiver theory given above. 

The general statement is now clear. Take a 5d quiver gauge theory, with  each $\SU(N)$ gauge node having effectively $2N$ flavors.
If  the quiver is a finite simply-laced Dynkin diagram of type $G$,
the instanton number currents enhance to  $G\times G$;
If  the quiver is an affine simply-laced Dynkin diagram of type $G$,
the instanton number currents enhance to  $\widehat{G\times G}$.

\section{Conclusions}\label{sec:conclusions}
In this paper we analyzed the one-instanton operators of 5d gauge theories with $\SU(N)$ gauge groups with hypermultiplets in the fundamental, adjoint, or bifundamental representations. 
We saw that a simple exercise in the treatment of fermionic zero modes gives rise to the expected patterns of symmetry enhancements. 

There are many areas to be further explored. One is to extend our analysis to include $\SU(N)$ gauge theories with other matter representations, such as antisymmetric or symmetric two-index tensor representations, and to consider other gauge groups, both classical and exceptional.  There should not be any essential difficulty to perform this generalization, since a one-instanton configuration in any group $G$ is always just an $\SU(2)$ one-instanton configuration embedded into $G$.  Our analysis of the $\SU$ quiver theory was by no means exhaustive, and it would be interesting to consider more general cases. 

It might be interesting to study instanton operators with higher instanton numbers. This will be significantly harder, however, since the instanton moduli space is much more complicated. Presumably, we will need to use the localization etc. to analyze it, and the method would become  equivalent to what has already been done in the literature in the study of the superconformal index of the 5d SCFTs.

Another direction is to study in more detail  the structure of the supermultiplets formed by operators in non-conformal 5d supersymmetric theories.  In this paper we relied on some heuristics based on the known supermultiplet structures of superconformal theories. The gauge theories in the infrared are however non-conformal, and we should analyze them as they deserve.  For example, in our analysis of $\SU(N_1)\times \SU(N_2)$ theory, we could not directly analyze the tensor product decomposition of the two copies of \eqref{instop} and the contribution \eqref{new}; we instead needed to import the knowledge gained by the analysis of the special case $\SU(2)^2$.  This is not an ideal situation. With a proper understanding of the supermultiplet structures of operators in non-conformal theories, we would be able to analyze this tensor product directly. 

We also assumed throughout in this paper that we only have to consider fermionic zero modes around the one-instanton configuration, and that the states with non-zero modes excited do not give broken current supermultiplets. This is at least plausible, since non-zero modes would likely produce descendant operators, but this is not at all a rigorous argument.  This needs to be better investigated. 

Finally, we assumed in this paper that the gauge theory we analyze is a mass deformation of a UV fixed point, either a five-dimensional one or a six-dimensional one compactified on $S^1$, and then studied what would be the enhanced symmetry in the ultraviolet. It would be desirable to understand the criterion to tell which 5d gauge theory has a UV completion. 

The author would like to come back to these questions in the future, but he will not have time in the next few months due to various duties in the university. He hopes that some of the readers get interested and make great progress in the meantime. 

\newpage

\section*{Acknowledgements}
The author  thanks the Yukawa Institute at Kyoto University, for inviting him to give a series of lectures in February 2015. The author, without much thinking, promised to give an introduction to  supersymmetric field theories in five and six dimensions.  It was after much thinking about how to organize the content that he came up with the arguments presented in this manuscript. 
The author also thanks Naoki Kiryu, Tatsuma Nishioka, Kantaro Ohmori, Masato Taki for discussions.
He also thanks Kazuya Yonekura and Gabi Zafrir for pointing out an error in the first version of the manuscript.
The work  is  supported in part by JSPS Grant-in-Aid for Scientific Research No. 25870159,
and in part by WPI Initiative, MEXT, Japan at IPMU, the University of Tokyo.

\bibliographystyle{ytphys}
\small\baselineskip=.9\baselineskip
\let\bbb\bibitem\def\bibitem{\itemsep1pt\bbb}
\bibliography{ref}
\end{document}